\documentclass[pre,groupedaddress,showkeys,showpacs,twocolumn]{revtex4}
\usepackage{amsmath}
\usepackage{graphics}
\usepackage{graphicx}
\usepackage{amsfonts}
\usepackage{amssymb}
\usepackage{dcolumn}
\usepackage{bm}
\usepackage{natbib}
\begin{document}
\title{Adhesive contact of model randomly rough rubber surfaces}
\author{Vito Acito}
\affiliation{Soft Matter Science and Engineering Laboratory (SIMM), UMR CNRS 7615, Ecole Sup\'erieure de Physique et Chimie Industrielles (ESPCI), Universit\'e Pierre et Marie Curie, Paris (UPMC), France}
\affiliation{Politecnico di Bari. Centre for Excellence in Computational Mechanics. Viale Gentile 182, 70126 Bari, Italy}
\author{Michele Ciavarella}
\affiliation{Politecnico di Bari. Centre for Excellence in Computational Mechanics. Viale Gentile 182, 70126 Bari, Italy}
\author{Alexis M. Prevost}
\affiliation{Laboratoire Jean Perrin, CNRS UMR~8237, Sorbonne Universit\'e, 4 Place Jussieu, 75005 Paris, France}
\author{Antoine Chateauminois}
\affiliation{Soft Matter Science and Engineering Laboratory (SIMM), UMR CNRS 7615, Ecole Sup\'erieure de Physique et Chimie Industrielles (ESPCI), Universit\'e Pierre et Marie Curie, Paris (UPMC), France}
\email[]{antoine.chateauminois@espci.fr}
\date{\today}
\begin{abstract}
	We study experimentally and theoretically the equilibrium adhesive contact between a smooth glass lens and a rough rubber surface textured with spherical microasperities with controlled height and spatial distributions. Measurements of the real contact area $A$ versus load $P$ are performed under compression by imaging the light transmitted at the microcontacts. $A(P)$ is found to be non-linear and to strongly depend on the standard deviation of the asperity height distribution. Experimental results are discussed in the light of a discrete version of Fuller and Tabor's (FT) original model (\textit{Proceedings of the Royal Society A} \textbf{345} (1975) 327), which allows to take into account the elastic coupling arising from both microasperities interactions and curvature of the glass lens. Our experimental data on microcontact size distributions are well captured by our discrete extended model. We show that the elastic coupling arising from the lens curvature has a significant contribution to the $A(P)$ relationship. Our discrete model also clearly shows that the adhesion-induced effect on $A$ remains significant even for vanishingly small pull-off forces. Last, at the local asperity length scale, our measurements show that the pressure dependence of the microcontacts density can be simply described by the original FT model.
\end{abstract}
\maketitle
\section{Introduction}
\label{intro}
Surface roughness has long been recognized as a major ingredient that cannot be ignored when considering adhesion and friction between macroscopic bodies. As emphasized by the pioneering work of Bowden and Tabor~\cite{Bowden1958}, a key component in the description of the tribological properties of rough bodies is the determination of the actual contact area formed by microasperity contacts of typical length scales distributed over orders of magnitude. This problem involves a still poorly understood interplay between the mechanical properties of the solid bodies, the geometry of the surfaces and adhesive forces. In the case of rubber-like materials, the effects of adhesive forces - such as Van der Waals forces- on contact deformation have long been recognized with the pioneering work of Johnson, Kendall and Roberts (JKR)~\cite{Johnson1971}. Based on the JKR model for the adhesion of a smooth sphere on a rigid flat, a first theoretical and experimental investigation of the effect of roughness on elastic contact with adhesion was carried out by Fuller and Tabor (FT)~\cite{fuller1975}. FT's model is based on Greenwood and Williamson's (GW)~\cite{greenwood1966} description of roughness which considers identical spherical asperities of radius $R$ whose summit heights is obtained from a Gaussian distribution with a standard deviation $\sigma$. Using JKR's solution for each asperity contact, FT obtained a solution for the rough adhesive contact problem, yielding an expression for the pull-off force which depends on the dimensionless parameter
\begin{equation}
\theta_{FT}=\frac{\sigma^{3/2}E^*}{R^{1/2}w}
\label{eq:adhesion_parameter}
\end{equation}
where $E^{\raisebox{0.7mm}{$ *$}}=E/(1-\nu^2)$ is the reduced Young's modulus ($\nu$ being Poisson's ratio), $R$ is the radius of the microasperities and $w$ is the adhesion energy. According to FT's model, the pull-off force decays exponentially with $\theta_{FT}$. In the same work, FT also conducted experiments where molded rubber spheres were put in contact with a Perspex flat with a random roughness obtained by bead-blasting or abrasion. However, such randomly rough surfaces are too complex to be properly amenable to the derived adhesive contact model. In addition, the lens curvature itself raises some issues. Even though FT's experiments show that lens curvature has a negligible role on the pull-off force, one can still question how it would affect local microcontact sizes and pressure spatial distributions.\\
\indent In this paper, we revisit FT's model for adhesive contacts of statistically rough surfaces in the light of experiments where we take advantage of recent micro\-milling techniques to engineer randomly rough silicone surfaces with prescribed distributions of both asperity lateral positions and heights. They consist of spherical asperities whose radii of curvature ($\approx 100~\mu$m) allow for an optical measurement of the spatial distribution of the microcontact areas when the patterned silicone substrate is indented by a smooth glass sphere. As compared to two previous studies~\cite{romero2013,yashima2015} using similar surfaces, the effects of adhesion were here enhanced by reducing both the elastic modulus and the small scale roughness of the asperities. Such surfaces are reminiscent of the regular arrays of spherical rubber caps investigated by Verneuil and coworkers~\cite{verneuil2007}. In this work, we get access to local quantities such as microcontact radii and contact pressures as a function of height distributions.\\
\indent To analyze our measurements within the framework of FT's model, we do focus on adhesive contact close to equilibrium conditions using a contact loading procedure for which viscoelastic dissipative effects are minimized. Adhesive effects at equilibrium are discussed by comparing experimental results for the load dependence of the real contact area to predictions of an adhesive multiasperity contact model which takes into account the effects of the elastic coupling arising from both lens curvature and microcontacts interactions. We also show that the use of textured surfaces allows for an accurate discussion of FT's adhesive contact model based on our investigation of the statistical distributions of microcontact radii and pressures. 
%
\section{Experimental details}
\label{sec:exp_details}
\begin{figure}[!ht]
	\includegraphics[width=0.9 \columnwidth]{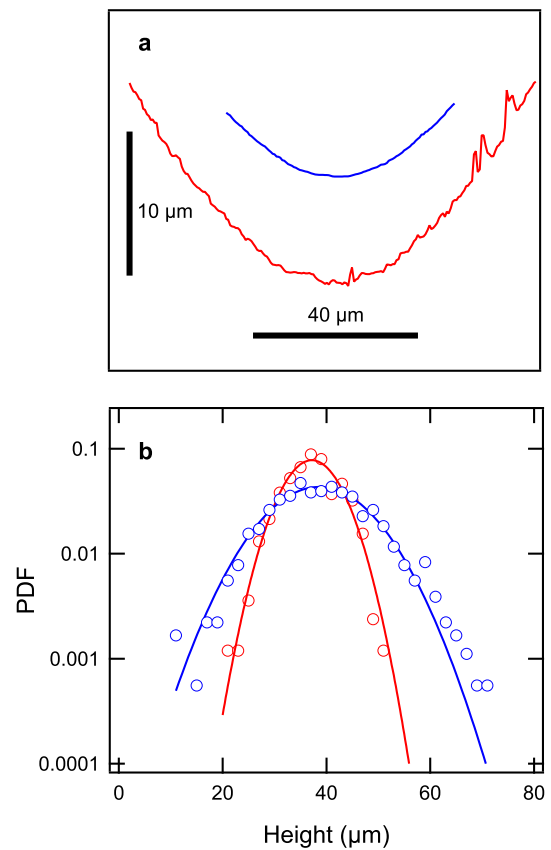}
	\caption{Topography of the micromilled PMMA mold. (a) Typical optical profilometry profiles of the spherical cavities along a diameter before (red, lower curve) and after (blue, upper curve) exposure to a saturated CHCl$_3$ vapor. (b) Probability density functions of the height distributions measured on both CHCl$_3$ vapor treated PMMA random molds using optical profilometry. Solid lines are Gaussian fits with $\sigma=5.13 \pm 0.19~\mu$m (red) and $\sigma=9.22 \pm 0.15~\mu$m (blue).}
	\label{fig:bumps_topography}
\end{figure}

\begin{figure}[!ht]
	\includegraphics[width=0.9 \columnwidth]{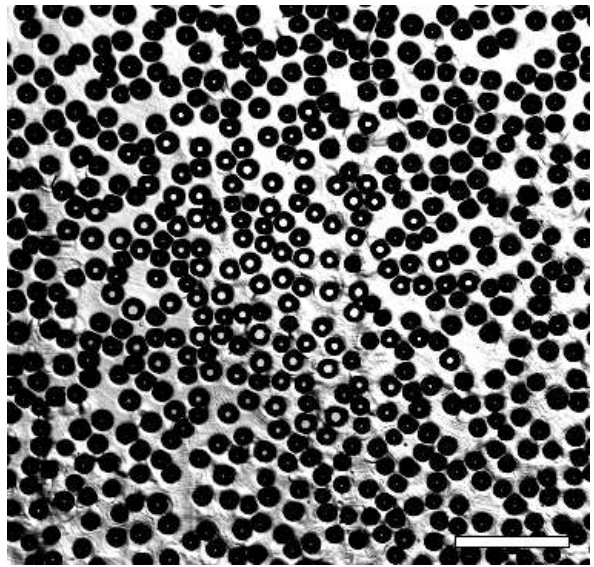}
	\caption{Image in transmission of a typical contact under compressive load ($P=12$~mN). Microcontacts appear as bright disks on top of the asperities. The white bar is 1~mm long.}
	\label{fig:contact_picture}
\end{figure}

Normal contact experiments are carried out between an optically smooth glass lens (radius of curvature 103.7 mm, BK7, Melles-Griot) and a nominally flat Poly\-DiMethylSiloxane (PDMS) slab decorated with micrometer sized spherical caps of equal radius of curvature, distributed randomly both spatially and in height. As detailed in~\cite{romero2013,yashima2015}, these patterned surfaces are obtained by crosslinking PDMS in Poly(MethylMethAcrylate) (PMMA) molds milled with a desktop CNC Mini-Mill machine (Minitech Machinary Corp., USA) using ball end mills of radius 100~$\mu$m. Prescribed random distribution of asperities were distributed over 1~cm$^2$ with a non-overlapping constraint and an asperity density of 2.10$^7$~m$^{-2}$. Three different patterns have been made on the same PMMA mold, one reference area covered with a square network of spherical caps having all the same height of 40~$\mu$m and two areas decorated with Gaussian height distributions of spherical caps with standard deviations $\sigma=5~\mu$m and $\sigma=10~\mu$m respectively. In addition, a fourth part of the PMMA mold has been kept smooth to allow measurements of the elastic modulus and the adhesion energy of the PDMS sample using a JKR contact configuration.\\
\indent As shown in Fig.~\ref{fig:bumps_topography}a, micromilled spherical caps on the PMMA mold inherently present a micrometric scale roughness which can induce a reduction of adhesion of the replicated patterned PDMS substrate. To minimize these effects, PMMA molds have thus been exposed to a saturated CHCl$_3$ vapor for 15~minutes. As a result of surface plasticization of the glassy acrylate polymer, surface tension effects were found to result in a smoothening of the surface of the spherical cavities of the mold (Fig.~\ref{fig:bumps_topography}a).
It was verified with optical profilometry that this treatment did not induce any change in the standard deviation of asperity height distribution. However, it resulted in a roughly 10\% increase in the radius of curvature of the asperities up to 110~$\mu$m which was accounted for in both data analysis and contact simulations. Figure~\ref{fig:bumps_topography}b shows for both random vapor treated PMMA molds, the Probability Density Functions (PDF) of the asperity height distributions measured over more than 400 asperities. PDFs are still well fitted by Gaussians with standard deviations very close to the prescribed ones, namely $\sigma=5.13 \pm 0.19~\mu$m and $\sigma=9.22 \pm 0.15~\mu$m.\\ 
\indent PDMS substrates are obtained by crosslinking at 70$^\circ$C for 48 hours a mixture of commercially available  Sylgard 184 and Sylgard 527 liquid silicones (Dow Corning, Midland, USA). As detailed in~\cite{palchesko2012}, mixing both silicones in various proportions allows tuning the elastic modulus in the [kPa--MPa] range. Here, Sylgard 527 and Sylgard 184 were mixed in a 0.75:0.25 weight ratio to achieve a Young's modulus of about 0.5~MPa. The elastic modulus $E$ and the adhesion energy $w$ of the flat PDMS substrate were obtained with standard JKR measurements with an optically smooth glass lens. These measurements were performed using the setup and the step-by-step contact loading procedure described below. They yielded $E=~0.51 \pm 0.11$~MPa and $w=~40.8 \pm 2.1$~mJ~m$^{-2}$.\\
\indent Normal contact experiments are carried out using a custom contact setup which is fully described elsewhere~\cite{piccardo2013}.  The glass lens indenter is fixed to a motorized vertical translation stage by means of a double cantilever beam with a stiffness of 290~N~m$^{-1}$. During loading, the normal load is determined with a 0.1~mN resolution from the measurement of the deflection of the cantilever using a high resolution optical sensor (Philtec D64-L). In addition, the vertical position of the lens is monitored with submicrometer accuracy using a laser sensor (Keyence LK-H057). The PDMS substrate is fixed to two crossed motorized linear translation stage allowing to change the relative position of the patterned surface with respect to the glass lens. Contact pictures are recorded through the transparent PDMS substrate using a zoom objective (Leica APO Z16) and a high resolution CMOS camera (SVS Vistek Exo, 2048 x 2048 pixels$^2$, 8 bits). Once illuminated in transmission with an LED light spot, microcontact appear as bright disks (Fig.~\ref{fig:contact_picture}) whose center and area (and radius $a_i$ for a given asperity $i$) can be measured with standard thresholding techniques~\cite{romero2013,yashima2015}. In the absence of any contact with the glass lens, images of the LED light source result in a small residual bright point at the apex of the microlenses. The radius of these bright spots thus sets to about 8~$\mu$m the detection threshold of microcontacts.\\
\indent As already mentioned in the introduction, our study focuses on adhesive contacts close to equilibrium. For glass--rubber single asperity contacts (including glass--PDMS), it is known that adhesive contact equilibrium is achieved more readily during contact loading than unloading~\cite{deruelle1998,Maugis1978}. We thus privileged a step-by-step contact loading procedure where the normal load $P$ is increased in discrete steps of 4~mN from 4~mN to 20~mN. At each loading step, it was however observed that the equilibrium adhesive state is not reached instantaneously. Indeed, the measured total contact area slowly increased over a duration that however never exceeded $\sim 1000$~s within the investigated range of normal loads. Consequently, in our experiments both the normal force and contact pictures are recorded after a dwell time of 1000~s following each loading step.
At the end of the loading procedure, the contact is unloaded at an imposed displacement rate between 2 and 20~$\mu$m~s$^{-1}$ from the maximum applied load of 20~mN. For each patterned surface, this procedure is repeated over 36 independent contact realizations by moving laterally the PDMS substrate in 0.5~mm long steps with respect to the fixed glass lens. Within the selected load range, no contact was found to occur in the smooth regions between spherical caps.
%
\section{Contact model}
\label{sec:contact_model}
Modeling the adhesive contact response was performed by extending to the adhesive case a discrete numerical model developed previously for non-adhesive randomly rough contacts~\cite{yashima2015}. As in FT's original model, this approach assumes that JKR theory holds at the microcontact length scale. Compared to FT's model, our model also takes into account to first-order elastic interactions arising from both the indenting lens curvature and microcontacts interactions. Following Ciavarella~\textit{et al.}~\cite{ciavarella2008}, one imposes a displacement which is sensitive to the effect of the spatial distribution of contact pressure in the neighboring asperities. More precisely, for each microasperity contact, a shift of the position of the deformable surface is introduced, which results from the vertical displacement caused by the neighboring ones. Accordingly, the indentation depth $\delta_i$ of the $i^{th}$ microasperity contact writes
\begin{equation}
\delta_i = \delta_i^0 + \sum_{j\neq i}^N\alpha_{ij}f(\delta_{j})\:,
\label{eq:delta}
\end{equation}
where $\delta_i^0>0$ is the indentation depth in the absence of any elastic coupling between microcontacts, and $\alpha_{ij}f(\delta_{j})$ are the elements of the interaction matrix. $\delta_i^0$ is a purely geometrical term simply given by the difference between the positions of the two undeformed surfaces for the prescribed indentation depth $\Delta$~\cite{yashima2015}. The sum in the rhs of eqn~(\ref{eq:delta}) represents the interaction term derived from JKR contact theory. We take for this term an asymptotic expansion of the JKR solution for the vertical displacement of the surface, instead of its exact expression. As detailed in the Supplementary Information, elements $\alpha_{ij}f(\delta_{j})$ of the interaction matrix thus read
\begin{equation}
[\alpha_{ij}f(\delta_{j})] = - \frac{4}{3 \pi} \frac{1}{r_{ij}} f(\delta_{j}) \:\:, i \neq  j\:,
\label{eq:matA}
\end{equation}
where $r_{ij}$ is the distance from the center of asperity $i$ to that of asperity $j$ and $R$ is the radius of curvature of the microasperities. The function $f(\delta_{j})$ corresponds to the JKR relationship between load and contact radius $a_i$ and writes
\begin{equation}
f(\delta_j)=\frac{3Ra_i(\delta_j)\delta_j-a_i^3(\delta_j)}{2R}
\end{equation}	
where $a_i(\delta_j)$ can be deduced from the numerical inversion of the explicit JKR expression for $\delta_i(a_i)$. For each value of the standard deviation $\sigma$ of asperity heights, calculations are repeated over 500 different contact realizations generated from Gaussian sets of asperity heights. Asperities are spatially distributed according to a uniform distribution with a non-overlap constraint. Consistently with the experiments, calculations are carried out by increasing step by step the gap between surfaces until the prescribed maximum load is achieved.\\
\indent Overall, this numerical model differs from FT's original model by 3 ingredients, (\textit{i}) non-vanishing elastic interactions, (\textit{ii}) a sphere-on-plane contact rather than plane-on-plane and (\textit{iii}) a discrete number of asperities. Note that ingredients (\textit{i}) and (\textit{ii}) can be tuned to get a discrete version of FT's model, by simply setting $[\alpha_{ij}f(\delta_{j})]=0$ in the simulations and by replacing the curved lens by a flat.\\	
%
\section{Results and discussion}

\subsection{Pull-off force and total contact area}
\label{sec:contact_area}
\begin{figure}
	\includegraphics[width=1 \columnwidth]{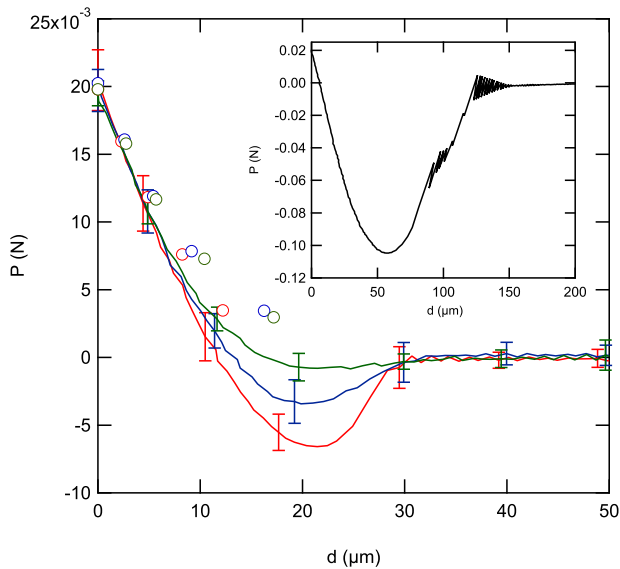}
	\caption{Normal load $P$ as a function of the displacement $d$ (solid lines) during unloading at an imposed displacement rate of 20~$\mu$m~s$^{-1}$, with $\sigma=0$ (red lowest curve), $\sigma=5~\mu$m (blue middle curve) and $\sigma=10~\mu$m (green upper curve). Open symbols (with the same color code) correspond to step-by-step contact loading with the same patterned surfaces. Inset: $P$ versus $d$ for the smooth PDMS substrate.}
	\label{fig:f_unload}
\end{figure}

\begin{figure}
	\includegraphics[width=1 \columnwidth]{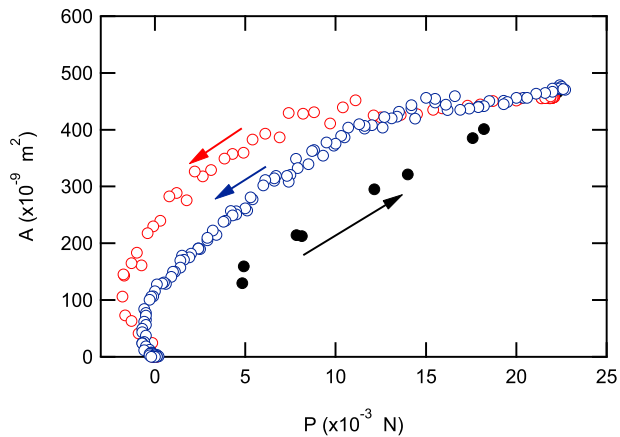}
	\caption{Total area of contact $A$ \textit{vs} normal load $P$ during a step-by-step contact loading (black disks) and unloadings at a constant imposed displacement rate of ~2~$\mu$m~s$^{-1}$ (blue circles) and ~20~$\mu$m~s$^{-1}$ (red circles). The patterned surface shown here has $\sigma=5~\mu$m. An adhesion hysteresis is observed between equilibrium loading and out-of equilibrium unloading.}
	\label{fig:a_p_unload}
\end{figure}

\begin{figure}[!ht]
	\includegraphics[width=1 \columnwidth]{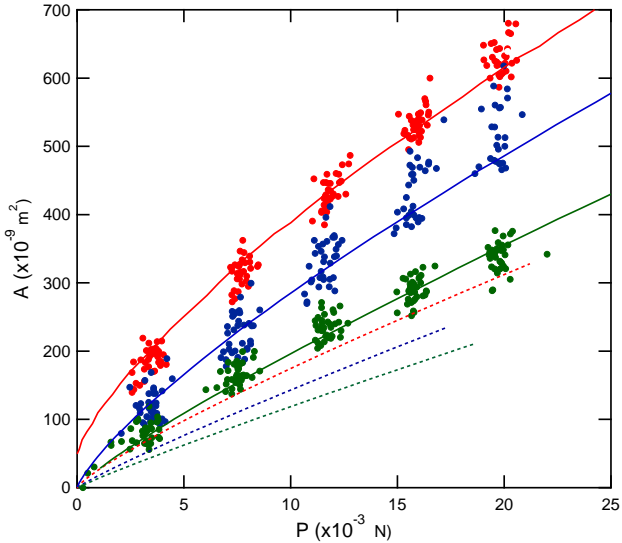}
	\caption{Total contact area $A$ as a function of the normal load $P$ with $\sigma=0$ (red disks), $\sigma=5~\mu$m (blue disks) and $\sigma=10~\mu$m (green disks). Solid and dotted lines (with the same color code) are the predictions of the adhesive and non-adhesive contact simulations, respectively.}
	\label{fig:total_cont_area}
\end{figure}

\begin{figure}[!ht]
	\includegraphics[width=1 \columnwidth]{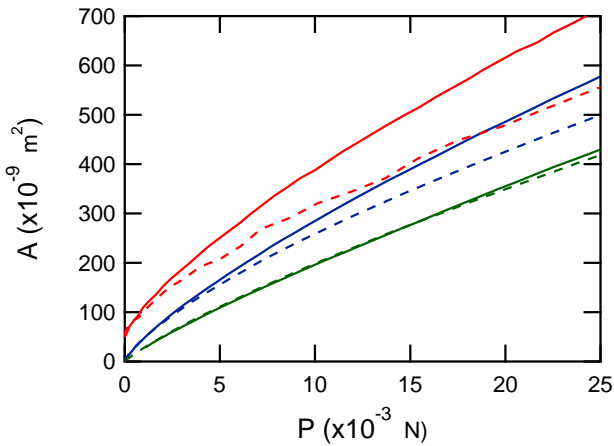}
	\caption{Calculated total contact area $A$ as a function of the normal load $P$ with $[\alpha_{ij}] \ne 0$ (solid lines) and $[\alpha_{ij}]=0$ (dotted lines) for a sphere-on-flat adhesive contact, with $\sigma=0$ (red curves), $\sigma=5~\mu$m (blue curves), $\sigma=10~\mu$m (green curves).}
	\label{fig:simul_interaction_sphere}
\end{figure}

\begin{figure}[!ht]
	\includegraphics[width=1 \columnwidth]{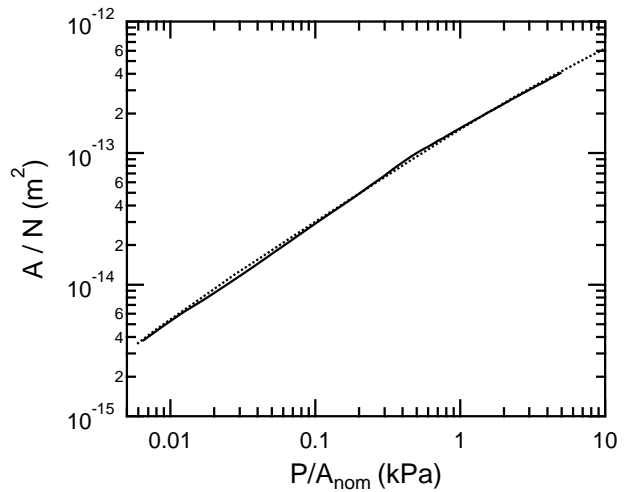}
	\caption{Log-log plot of the calculated total area of contact $A$ normalized by the number of asperities $N$ as a function of the applied load $P$ divided by the nominal apparent contact area $A_{nom}$ for nominally flat surfaces with $\sigma=5~\mu$m. Calculations are carried out with (solid lines) and without (dotted lines) elastic interactions.}
	\label{fig:simul_interaction_flat}
\end{figure}

We have first considered how integrated quantities, \textit{i.e.} the pull-off force $P_c$ and the total area of contact $A$  depend on both the loading parameters and the topography characteristics of the PDMS substrates. Figure~\ref{fig:f_unload} shows as solid lines the relationship between normal load $P$ and vertical displacement $d$ during unloading at imposed displacement rate after a dwell time of 1000~s at $P=$20~mN. Each curve is an average over 36 contact realizations with the standard deviation indicated by the error bars. For the constant height asperity pattern ($\sigma=0$), the measured average pull-off force is $P_c=6.6 \pm 1.3$~mN. For $\sigma=5 \mu$m and $\sigma=10 \mu$m, the pull-off force is reduced to $P_c=3.4 \pm 1.6$~mN and $P_c=0.8 \pm 1.0$~mN, respectively. As a reference, the contact unloading response of the smooth glass/PDMS contact is also shown in the inset of Fig.~\ref{fig:f_unload}, yielding a pull-off force of $118 \pm 5$~mN. It turns out that roughness induces a decrease in the pull-off force by about two orders of magnitude. Such a strong decrease can be discussed in the light of the adhesion parameter $\theta_{FT}$ introduced by FT in their model (Eq.~\ref{eq:adhesion_parameter}). Physically, $\theta_{FT}$ represents the ratio of the elastic force needed to push one microasperity to a depth $\sigma$ to the adhesive pull-off force experienced by the same microasperity. In other words, it expresses the competition between compressive forces exerted by the higher asperities and the adhesive force between the lower asperities with respect to the mean nominal interface: when the roughness and hence the adhesion parameter increases, the adhesion falls to small values. Taking into account the elastic and adhesive properties of the PDMS substrate ($E=$~0.51~MPa and $w=$~40.8~mJ~m$^{-2}$) and the radius of the microasperities, one gets $\theta_{FT}$~=13.7 and 37.7 for $\sigma=$~5 and 10~$\mu$m, respectively. For such high $\theta_{FT}$ values, FT predict that the ratio of the pull-off force of the rough contact $P_c$ to the one measured with the same spherical surface and a smooth PDMS flat $P_c^s$, vanishes (\textit{cf} Fig.~6 in~\cite{fuller1975}). Consistently with this prediction, we find experimentally that the ratio $P_c/P_c^s$ is very low ($3.10^{-2}$ and $6.10^{-3}$ for  $\sigma=5 \mu$m and $\sigma=10 \mu$m, respectively).\\
\indent Noticeabily, the measured values of the pull-off force exhibit a rate dependence which is indicative of dissipative effects. When the unloading velocity is decreased to $v=2~\mu$m~s$^{-1}$, the pull-off force is reduced to $2.1~\pm~0.3$~mN for the random substrate with $\sigma=5$~$\mu$m (results not shown). Conversely, increasing $v$ to $200~\mu$m~s$^{-1}$ yields an enhanced pull-off force of $9.3 \pm 0.6$~mN. Note that some hysteresis, consistent with the occurrence of dissipative effects, is also evidenced when step-by-step loading and constant rate unloading curves are compared (\textit{see} open symbols \textit{versus} solid lines on Fig.~\ref{fig:f_unload}).\\
\indent This adhesion hysteresis due to viscoelastic loading can be best seen when plotting the total contact area $A$, obtained by summing up all individual microcontact areas, \textit{versus} $P$, for both step-by-step loading and constant rate unloading. This is shown in Fig.~\ref{fig:a_p_unload} with the example of the patterned surface with $\sigma=5~\mu$m. A strong hysteresis is observed, that is enhanced with the unloading rate, consistently with the assumption of viscoelastic effects.\\
\indent In the case of single-asperity adhesive contacts with rubbers, dissipative viscoelastic effects have long been reported to enhance adhesion (\textit{see e.g.}~\cite{barquins1977,deruelle1998,barthel2009}). Here, they are induced at the length scale of microasperity contacts. For the Sylgard silicone used in our experiments, viscoelastic dissipation may especially arise from the presence of free chains~\cite{hourlier2017} and dangling bonds~\cite{amoroux2003} within the polymer network. Such viscoelastic effects are obviously not taken into account in the FT's theory which is based on an equilibrium description of adhesive contacts. In what follows, we now focus on adhesive equilibrium during contact loading only.\\	
\indent Figure~\ref{fig:total_cont_area} shows $A(P)$ for all three asperity height distributions. As detailed in Section~\ref{sec:exp_details}, we recall that this data was obtained close to adhesive equilibrium by enforcing a dwell time subsequent to each load increment. For each distribution, data points correspond to 36 contact realizations. For a given load $P$, increasing $\sigma$ results in a clear decrease of $A$. For a given $\sigma$, $A(P)$ also shows some non-linearity which departs from the observed behavior with Hertzian contacts with similar surfaces~\cite{yashima2015}.\\
To further assess the effects of adhesion on the measured $A(P)$, contact simulations were carried out using our extended discrete model with either $w=0$ or $w=40$~mJ~m$^{-2}$. As shown with the solid lines in Fig.~\ref{fig:total_cont_area}, taking into account the elastic and adhesive properties of the PDMS substrate provides an accurate description of the experimental $A(P)$ within experimental errors. Interestingly, comparison with non-adhesive simulations (\textit{see} the dotted lines in Fig.~\ref{fig:total_cont_area}) shows that adhesion has a significant contribution to the load dependence of the contact area. The latter is indeed enhanced by a factor of roughly two depending on $\sigma$. Interestingly, such enhancement of the contact area is observed within a range of $\theta_{FT}$ values for which FT's model predicts a vanishing small effect of adhesion on the pull-off force. This experimental observation is consistent with the findings of theoretical adhesive rough contact models (\textit{see e.g.}~\cite{persson2008a}) which also predict an enhancement of the contact area with adhesion in the limiting case where no pull-off is can be detected.\\	
In addition, it is worthwhile noticing that the effects of adhesion on $A(P)$ decrease with $\sigma$, consistently with FT's model predictions.\\
Another issue is the contribution of the elastic coupling to the $A(P)$ relationship. This point is directly addressed in the contact simulations by turning on or off the elastic interaction term $\alpha_{ij}f(\delta_{j})$. For the sphere-on-flat contacts under consideration, both long range elastic interactions due to lens curvature and short-range interactions between neighboring microasperity contacts can be involved. In an attempt to quantify both effects, calculations were carried out either for a sphere-on-flat contact situation or for contacts between nominally flat surfaces. As shown in Fig.~\ref{fig:simul_interaction_sphere}, the effect of the elastic interactions for sphere-on-flat contacts is clearly to decrease the real contact area at a given load. If only short range interactions were involved, the reverse trend would be expected as the effects of such interactions is to induce a shift of the position of the deformable surface which in turn results in a reduction of actual microasperity indentation depths. 
Indeed, the fact that simulations for nominally flat surfaces with the interaction term either turned on or off yields the same behavior (Fig.~\ref{fig:simul_interaction_flat}) shows that, for the microasperity distributions under consideration, short range elastic coupling between neighboring microcontacts are unimportant.
As a consequence, it can be argued that elastic interactions mostly arise from long range effects due to the lens curvature.
\subsection{Microcontact radius and contact pressure distributions}
\begin{figure} [!ht]
	\includegraphics[width=1 \columnwidth]{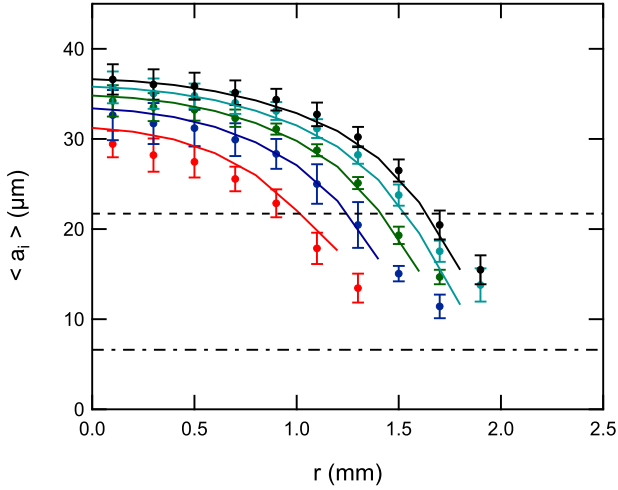}
	\caption{Radial profiles of average microcontact radius $<a_i>$. Red: $P=4$~mN, blue: $P=8$~mN, green: $P=12$~mN, light blue: $P=16$~mN, black: $P=20$~mN. Solid lines correspond to adhesive contact simulations. The dot-dashed and dashed lines correspond to the theoretical values of pull-off (at fixed grips) and zero load, respectively. Error bars correspond to standard deviations over 36 contact realizations.}
	\label{fig:a_p_s0}
\end{figure}

\begin{figure} [!ht]
	\includegraphics[width=1 \columnwidth]{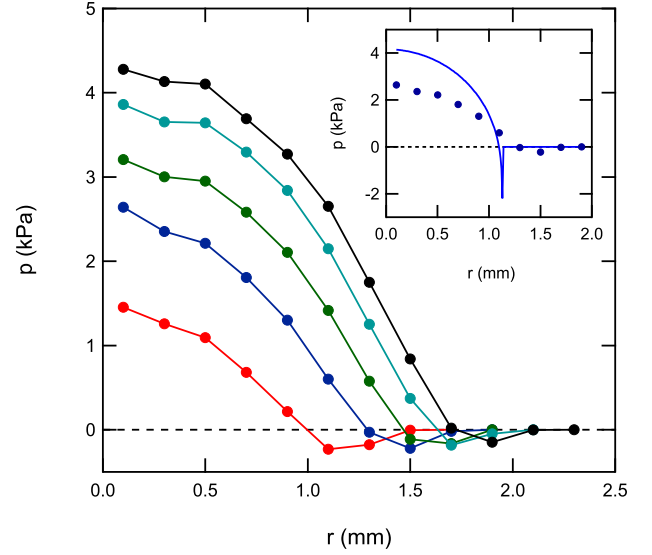}
	\caption{Radial profiles of contact pressure $p(r)$ for the patterned surface with $\sigma=0$ at $P=4$~mN (red curve), $P=8$~mN (blue curve), $P=12$~mN (green curve), $P=16$~mN (light blue), and $P=20$~mN (black curve). The dotted line corresponds to $p=0$. Inset: attempt to fit the experimental contact pressure data at $P=8$~mN to JKR's model with the adhesion energy $w$ as a fitting parameter. The fit yields $w=1$~mN.}
	\label{fig:cont_press_s0}
\end{figure}

\begin{figure} [!ht]
	\includegraphics[width=1 \columnwidth]{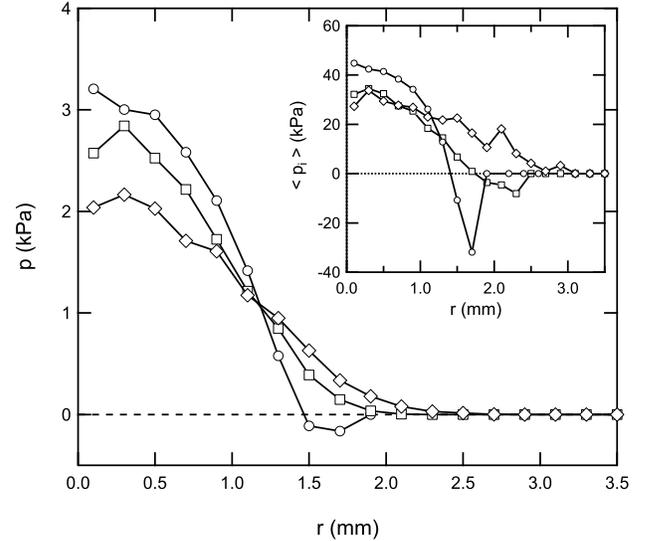}
	\caption{Radial profiles of contact pressure $p(r)$ at $P=$12~mN for $\sigma=0$ ($\circ$), $\sigma=5~\mu$m ($\square$) and $\sigma=10~\mu$m ($\diamond$). Inset: corresponding average microcontact pressure $<p_i>$ profiles for the same conditions. }
	\label{fig:press_12mn}
\end{figure}

\begin{figure} [!ht]
	\includegraphics[width=1 \columnwidth]{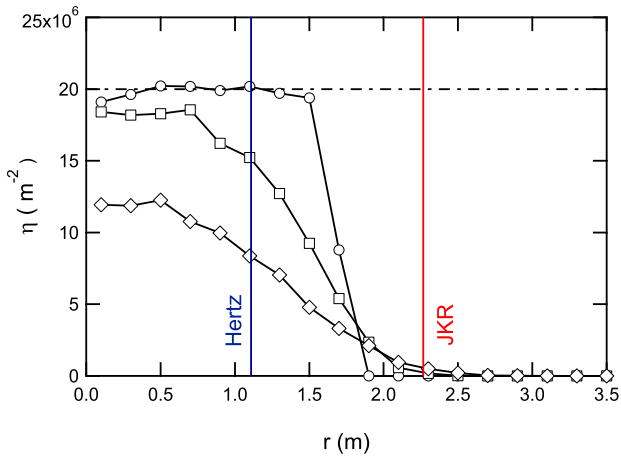}
	\caption{Radial profiles of microcontact density $\eta(r)$ at $P=12$~mN for $\sigma=0$ ($\circ$), $\sigma=5 \mu$m ($\square$) and $\sigma=10 \mu$m ($\diamond$). The vertical blue (\textit{resp.} red) line marks the theoretical value of the contact radius according to Hertz (\textit{resp.} JKR)'s theories with $E=0.51$~MPa and $w=40.8$0~mJ~m$^{-2}$. The horizontal dashed dotted line corresponds to the average microasperity density.}
	\label{fig:prof_eta_12mn}
\end{figure}

\begin{figure} [!ht]
	\includegraphics[width=1 \columnwidth]{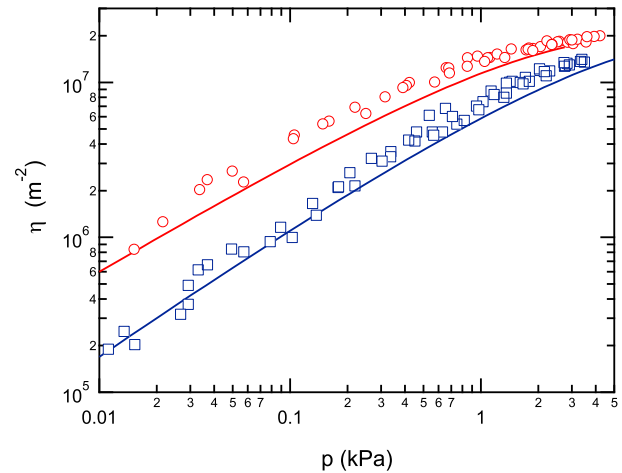}
	\caption{Log-log plot of the dependence of the average microcontact density $\eta$ with the local contact pressure $p$. Experimental data points for $\sigma=0$ (\textit{resp.} $\sigma=10~\mu$m) are shown with the red circles (\textit{resp.} blue squares). Solid lines are predictions of FT's model.}
	\label{fig:eta_p}
\end{figure}

We have then considered local quantities, \textit{i.e.} the radial distribution of microcontact radii $a_i$, the asperity density $\eta$ and the contact pressure $p$. Figure~\ref{fig:a_p_s0} shows the angularly averaged radial profiles of $a_i$, $<a_i>(r)$ for the regular array of microasperities ($\sigma=0$) measured at different applied normal loads $P$. Here, $r$ is the radial coordinate from the center of the apparent contact, taken as the barycenter of all microasperity contacts. In Fig.~\ref{fig:a_p_s0}, error bars correspond to the measured standard deviation over 36 different contact realizations. Also shown on Fig.~\ref{fig:a_p_s0} with the solid lines are the predictions of our discrete FT model, which are found to be in good agreement with $<a_i>(r)$ measured experimentally.\\
\indent Measured values of the microcontact radii can be compared to the theoretical values $a_0=(6 \pi w R^2/K)^{1/3}$ and $a_{min}=(\pi wR^2/6K)^{1/3}$ ($K=4E^{*}/3$ being an elastic constant) that correspond, within the framework of JKR's model, to zero load and pull-off at fixed grips respectively. With the elastic and adhesive properties of the considered silicone substrate, $a_{min}=6.58~\mu$m, a value that is close to the detection threshold of our measurements (about 8~$\mu$m) and $a_0=21.7~\mu$m. Both values are marked in Fig.~\ref{fig:a_p_s0} with the dashed dotted and dotted lines, respectively. As can clearly be seen in Fig.~\ref{fig:a_p_s0}, asperities lying at the periphery of the contact have microcontact radii that systematically fall between $a_{min}$ and $a_0$. Accordingly, a significant portion of the microasperity contacts are under tensile load at the periphery of the apparent contact.\\
\indent Radial contact pressure profiles $p(r)$ were obtained by summing up local forces $P_i$ exerted on all microcontacts located within an annulus of width $dr=0.2$~mm and radius $r$. For each microcontact, the value of $P_i$ was determined from the measured contact radius $a_i$ using the JKR relationship. Figure~\ref{fig:cont_press_s0} shows $p(r)$ for $\sigma=0$ and increasing normal loads $P$. Such $p(r)$ profiles are amenable to a direct comparison with JKR's predictions for a smooth sphere-on-flat contact. Tentative fits using JKR's model were thus performed taking the adhesion energy $w$ as the only fitting parameter. As shown in the inset of Fig.~\ref{fig:cont_press_s0} with the example of $P=8$~mN, deviations between theory and experiments are clearly evidenced. Recently, Degrandi-Contraires and coworkers~\cite{degrandi2013} have studied the adhesive contact between a flat PDMS decorated with cylindrical pillars of equal heights and a spherical indenter. In their analysis, they measured the apparent contact area as a function of the normal load. Using JKR's theory, they deduced from their measurements an effective adhesion energy attributed to the inherent roughness of the substrate. However, our results in Fig.~\ref{fig:cont_press_s0} clearly show the limits of applying to our rough contact interfaces such an approach solely based on JKR's theory with an effective adhesion energy. In other words, this means that for the patterned surfaces used in our experiments, the separation of length scales between the microasperity contacts and the macroscopic contact region is insufficient to allow for the use of a JKR model based approach with an effective adhesion energy.\\
\indent Figure~\ref{fig:press_12mn} shows $p(r)$ for surfaces with increasing $\sigma$, with the example of $P=12$~mN. Clearly, increasing $\sigma$ results in a decrease of the magnitude of tensile contact pressures at the contact periphery. For both $\sigma=5~\mu$m and $\sigma=10~\mu$m, such tensile contact pressures are barely detected. Average mean contact pressure measurements $<p_i=P_i/\pi a_i^2>$ carried out at the scale of individual micro-asperity contacts (inset of Fig.~\ref{fig:press_12mn}) shows that some asperities at the periphery of the apparent contact are still under a slight ($<p_i> \approx -8$~kPa) net tensile stress for $\sigma=5~\mu$m, while all micro-contacts are under compressive load for $\sigma=10~\mu$m.\\ 
\indent The dependence of adhesion on the standard deviation of asperity height distributions can be further analyzed in the light of measurements of the extension of the apparent contact area. The later can be determined quantitatively from the radial profiles of the microcontact density $\eta(r)$. Such profiles are reported in Fig.~\ref{fig:prof_eta_12mn} for $P=12$~mN and for $\sigma=0$, 5 and 10~$\mu$m. The vertical blue and red lines correspond to the calculated contact radius for non-adhesive and adhesive contacts of the spherical probe with the smooth PDMS substrate, respectively. From these profiles, one can determine an apparent contact radius $a_{app}$ by arbitrarily setting a threshold on $\eta$. Taking this threshold equal to 5~10$^{3}$~m$^{-2}$ yields $a_{app}=1.9, 2.5$ and 2.9~mm for $\sigma=0, 5$ and 10~$\mu$m, respectively. For $\sigma=0$, the effective contact radius is clearly lower than the one obtained for a smooth contact thus reflecting the decrease in adhesion induced by roughness. When $\sigma$ is increased, it turns out that the effective contact radius is increased and can even become close to that of the smooth JKR contact. Such feature can be explained by considering the occurrence of two opposite effects. On one side, the roughness induced reduction in adhesion leads to a decrease in the apparent contact size. On the other side, distributing microasperities along the vertical direction extends the size of the effective contact zone with respect to that achieved with all asperities at the same height (\textit{i.e.} for $\sigma=0$). New microcontacts are thus added in an annular region around the contact where the glass lens/PDMS separation is comparable to the surface roughness $\sigma$. Similar effects have already been reported for non-adhesive contacts~\cite{greenwood1967,yashima2015} and the extent of the annular region has been shown to scale as $R^{5/9}\sigma^{2/3}$. Here, a similar effect could be invoked in the case of adhesive contacts to explain the extension of the contact at a given applied load when $\sigma$ is increased.\\
\indent We now examine how the microcontact density $\eta$ depends on the local contact pressure $p$. Figure~\ref{fig:eta_p} shows, for all considered contact loads, a log-log plot of $\eta$ for both $\sigma=5$ and 10~$\mu$m as a function of $p(r)$. Also shown in this graph are solid lines that correspond to the theoretical predictions of the original FT's model (\textit{see} Appendix A). Clearly, FT's model is able to capture satisfactorily the measured $\eta(p)$ over the investigated pressure range with no need to incorporate any elastic interactions between microasperity contacts. Yet, one can note that FT's model slightly underestimates the measured microcontact densities. A possible explanation for this discrepancy could be that the JKR theory embedded in FT's model neglects the "jumping-on" phenomenon. The latter occurs when the distance between interacting surfaces becomes of the order of magnitude of the range of surface forces. As a result, FT's model should underestimate the microasperity density as well as the total area of contact. As detailed in a recent paper by Greenwood~\cite{greenwood2017}, FT's model can be extended to take into account such effects. However, comparison between experiments and contact simulations suggests that such effects are limited for the surfaces considered here.\\
\section{Conclusion}
In this work, we have investigated the adhesive contact between a smooth glass lens and a silicone surface decorated with prescribed distributions of spherical microasperities differing in their standard deviations of asperity heights. Experimental data were analyzed in the light of a discrete adhesive contact model allowing to account for elastic coupling effects arising from lens curvature and interactions between neighboring asperities. As in the original FT's theory, the model developed in this work postulates a JKR constitutive law at the length scale of individual microasperity contacts. Calculations are carried out under the assumption that the elastic and adhesive properties measured in a single asperity contact with a lens of radius $R=103.7$~mm can be transposed to the micrometer scale in order to describe the adhesive contacts between the soft micro-asperities and the glass surface. Our model allows to describe satisfactorily the experimental relationship between contact area and load for the various height distributions under consideration. Noticeably, it also indicates that adhesion increases significantly the contact area even if roughness results in a vanishingly small pull-off force as compared to a smooth contact interface. Such an effect would obviously be relevant to the description of the frictional properties of the rough contact interface which remains to be done. From a measurement of the microcontacts areas, we were able to determine the contact pressure distribution within the rough contact, which preserves one of the main feature of smooth JKR contacts, \textit{i.e.} the occurrence of tensile stresses at the contact periphery. We also showed that the original FT's model allows for an accurate description of the dependence of the microcontacts density on contact pressure, without any need to take into account elastic coupling between neighboring asperities. Such a conclusion holds for the considered surface topographies with asperities a single length scale ($R=100~\mu$m). However, theoretical calculations with more complex fractal surfaces indicate that the elastic coupling becomes important when surface topography is enriched with asperities at various length scales~\cite{Perrson2006}. This issue could be adressed experimentally using more sophisticated patterned surfaces with hierarchical distributions of microasperities. For instance, as a first step, this could be done by removing the non-overlaping constraint used in the design of our patterned surfaces.\\
\section*{Appendix: FT model}
In FT's model, the summits of the spherical asperities have a Gaussian height distribution given by $\phi(z)=\frac{1}{\sigma \sqrt{2\pi}}exp\left(-z^2/2\sigma^2\right)$. We consider here non-dimensional heights $\zeta \equiv z/\sigma$ so that the height distribution becomes $\phi(z)=\frac{1}{ \sqrt{2\pi}}exp\left(-\zeta^2/2\right)$. We follow the normalization suggested by Greenwood~\cite{greenwood2017} and define the normalized load $\overline{P}$, indentation depth $\overline{\delta}$ and contact radius  $\overline{a}$ as
\begin{equation}
\overline{P}=\frac{P}{Rw};\:\:	\overline{\delta}=\beta \frac{\delta}{R};\:\:\overline{a}= \beta\frac{a}{R}
\end{equation}
with
\begin{equation}
\beta \equiv \left(\frac{E*R}{w}\right)^{1/3}
\end{equation}	
Using this normalization, JKR equations now write
\begin{equation}
\overline{\delta}=\overline{a}^2-\sqrt{2\pi \overline{a}};\:\:\overline{P}= 4/3 \overline{a}^3 - \sqrt{8\pi \overline{a}^3}
\end{equation}
during loading. Assuming that the asperities do not jump into contact, the average load can be expressed as a function of the normalized mean plane separation $h \equiv d/\sigma$ as
\begin{equation}
\overline{P}(h)=\frac{N Z}{\sqrt{2\pi}} \int_0^\infty f\left(\overline{\delta}\right) \exp\left[-\left(h+Z\overline{\delta}\right)^2/2\right] d\overline{\delta} 
\end{equation}	
where $Z \equiv \left(Rw^2/E^{*2}\sigma^3\right)^{1/3}=4/3\pi^{2/3}\theta_{FT}$ is FT's adhesion parameter without the numerical factor $(3/4)/\pi^{2/3}$. As detailed in Appendix~2 of \cite{greenwood2017}, the above expression can be turned to an explicit integral which can readily be evaluated numerically. Similarly, one can write the microcontact density $\eta$ as
\begin{equation}
\eta(h)=\eta_0 \int_h^{\infty} \phi(\zeta)d\zeta=-\eta_0/2  \left[\mbox{erf}\left(h/\sqrt(2)  \right)-1\right]
\end{equation}	
where $\eta_0$ is the microasperity density.\\
\begin{acknowledgements}
	The authors wish to thank Guido Violano for his kind help in the obtention of some of the experimental data.
\end{acknowledgements}
%

%
%
%
\end{document}


\begin{center}
\bigskip{ {Acito \textit{et al} Adhesive contact of model randomly rough rubber surfaces}}.\\

\Large {Supplementary Information}\\

\bigskip{ {Derivation and implementation of the multiasperity adhesive contact model}}\\

\end{center}

\section{Derivation of the model}

\begin{figure} [ht]
	\centering
	\includegraphics[width=0.7\linewidth]{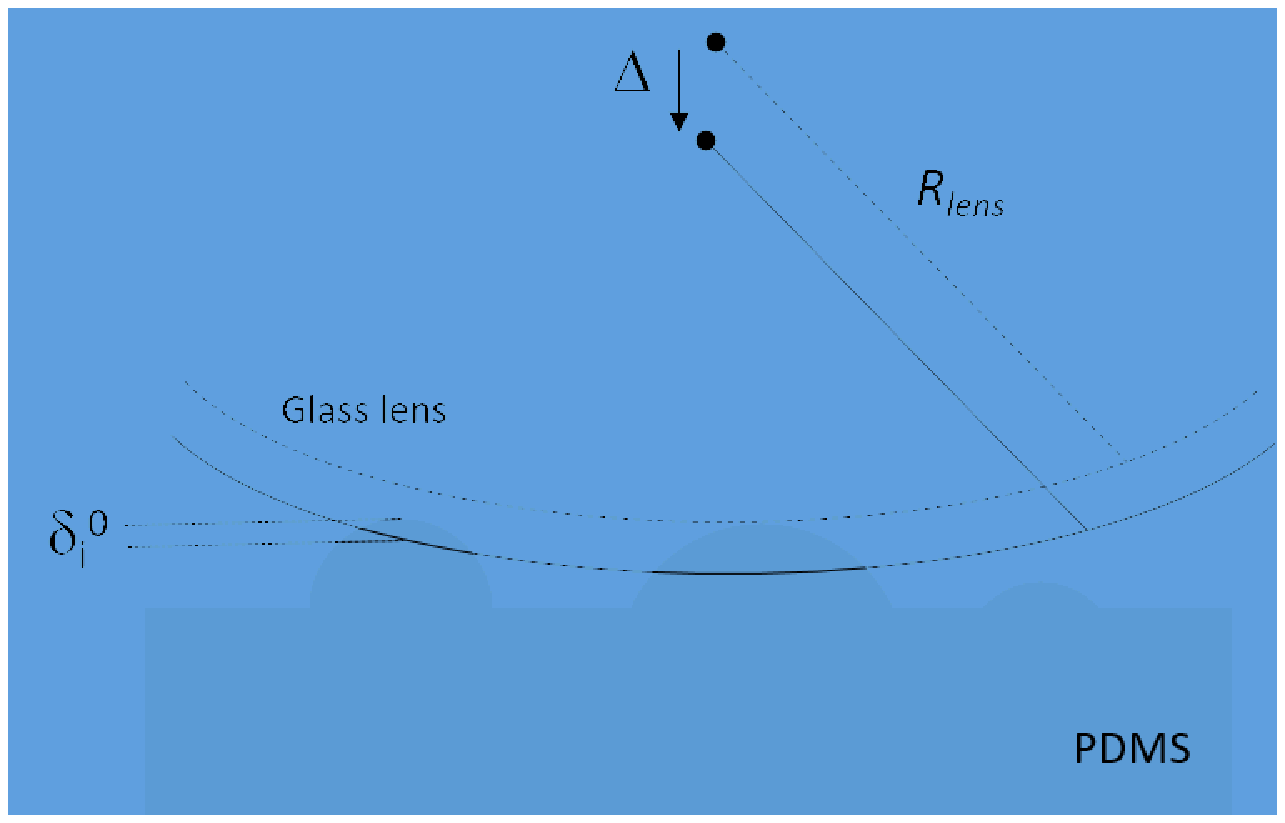}
	\caption{Sketch of the geometric configuration for the indentation of the patterned surface by a smooth rigid lens. $\Delta$ is the prescribed indentation depth taking as a reference for the vertical position of the indenting sphere the altitude at which the smooth surface is touching the uppermost asperity.}
	\label{fig:fig_si1}
\end{figure}

\begin{figure} [ht]
	\centering
	\includegraphics[width=0.7\linewidth]{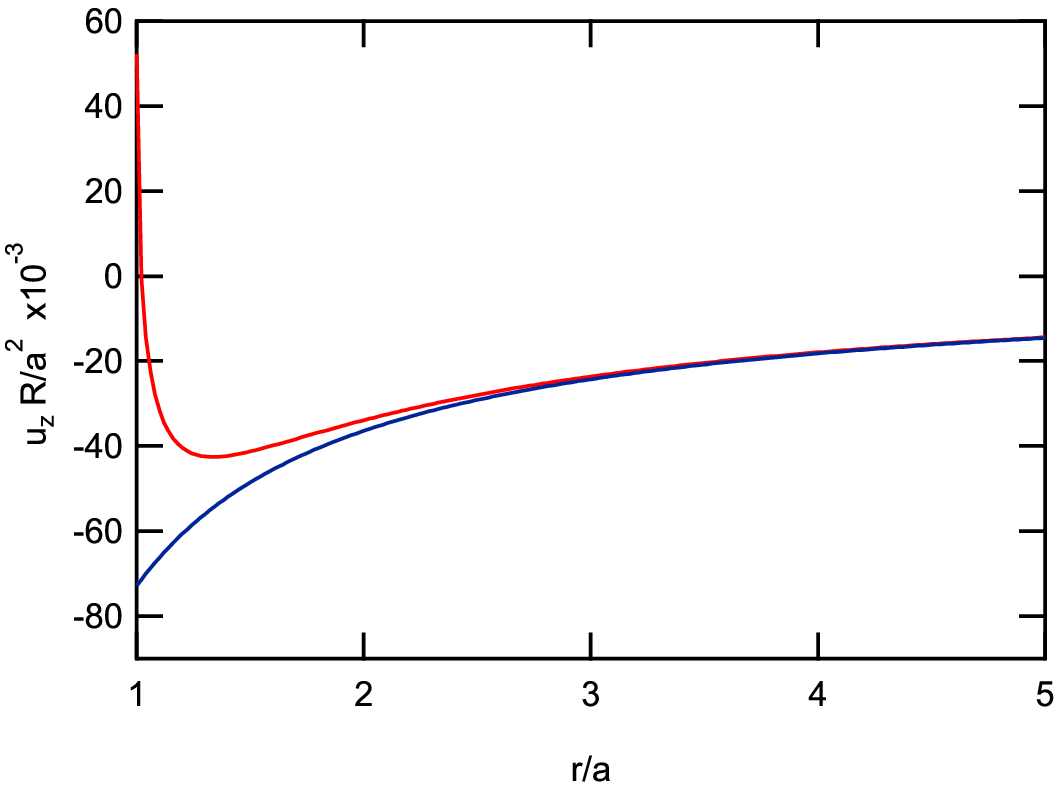}
	\caption{Normalized vertical displacement of the surface $u_z R/a^2$ calculated using the exact JKR relationship (red curve) and the asymptotic expression given in Eq.~\ref{eq:delta} (blue curve) for $P/(3\pi w R)=0.5$ ($R$ is the radius of the sphere, $r$ is the radial coordinate and $a$ is the contact radius).}
	\label{fig:fig_si2}
\end{figure}

We consider the indentation of a nominally flat elastic plane decorated with a distribution of spherical caps by a rigid, smooth and spherical probe. The rough surface is described by a collection of $N$ spherical asperities having all the same radius of curvature $R$ and prescribed positions $(x_i,y_i,z_i)$. Locally, the indentation of each asperity is assumed to obey JKR's theory. Elastic interactions between micro-contacts are accounted for by introducing a shift of the position of the deformable surface seen by each asperity due to the deflection caused by the neighboring ones. Accordingly, the actual indentation $\delta_i$ of the $i^{th}$ asperity is thus given by
\begin{equation}
\delta_i = \delta_i^0 + \sum_{j\neq i}^N\alpha_{ij}(\delta_j) 
\label{eq:delta}
\end{equation}
where $\delta_i^0>0$ is the indentation depth in the absence of any elastic coupling between microcontacts. As shown in Fig.~\ref{fig:fig_si1}, $\delta_i^0$ is a purely geometrical term simply given by the difference between the position of the two undeformed surfaces for the prescribed indentation depth $\Delta$.
The sum in the rhs of Eq.~\ref{eq:delta} represents the interaction term derived from JKR's theory. Instead of providing its exact expression, we derive $\alpha_{ij}$ from an asymptotic expansion that is based on the Green's tensor~\cite{landau1986} for a point loading on an elastic incompressible half-plane. It writes
\begin{equation}
\left(\alpha_{ij}(\delta_j) \right) = - \frac{3}{4 \pi E}\frac{1}{r_{ij}}F(\delta_j) \:\:, i \neq  j
\end{equation}
where $r_{ij}$ is the distance separating asperities $i$ and $j$ and where $F(\delta_j)$ is the normal indentation load of the $j^{th}$ asperity as obtained from JKR's theory
\begin{equation}
F(\delta_j)=\frac{3RK a_j(\delta_j)\delta_j-Ka_j^3(\delta_j)}{2R}
\label{eq:JKR}
\end{equation}	
where $K=16/9E$ is an elastic constant and $a_j(\delta_j)$ is the contact radius of the $j^{th}$ asperity. The latter can be determined by numerically inverting the explicit $\delta(a)$ JKR's relationship given by
\begin{equation}
\delta=\frac{a^2}{R}-\sqrt{\frac{8 \pi a w}{3K}} 
\end{equation}
where $w$ is the adhesion energy. From Eq~\ref{eq:JKR}, the interaction term $\alpha_{ij}(\delta_j)$ can be rewritten as\\
\begin{equation}
\left(\alpha_{ij}(\delta_j) \right)= - \frac{4}{3 \pi}\frac{1}{r_{ij}}f(\delta_j) \:\:, i \neq  j
\end{equation}
with
\begin{equation}
f(\delta_j)=\frac{3R a_j(\delta_j)\delta_j-a_j^3(\delta_j)}{2R}
\end{equation}
This asymptotic expression of the interaction term is expected to be valid as long as the distance $r_{ij}$ between two neighboring microcontacts is large with respect to their respective contact radii. From a comparison with the exact JKR expression~\cite{maugis1999} for vertical displacement of the free surface (Fig.~\ref{fig:fig_si2}), we found that the asymptotic expression provides a very accurate value of the surface displacement since $r_{ij} > 3a_{i,j}$, which is always verified experimentally.\\
\section{Numerical implementation}
Writing $V_i=z_i - d$, with $d$ being the approach distance of the two surfaces at the location of the $i^{th}$ asperity, the contact problem is the solution of the following set of non-linear equations
\begin{equation}
\mathbf{F\left(\delta \right)}=\delta - V - \mathbf{A} .\varphi \left(\delta\right)=0
\end{equation}
with $ \mathbf{A}=\left(\alpha_{ij}(\delta_j) \right)$ and
\begin{equation}
\varphi(u)=f(u) \:\: u\geqslant 0\:\:;\:\: \varphi(u)=0 \:\: u<0\\ 
\end{equation}
This set of equations is solved using a Newton-Raphson algorithm as follows
\begin{equation}
\delta_{n+1}=\delta_{n}- \mathbf{J}^{-1}_F \left( \delta_n\right) . \mathbf{F}\left( \delta_n\right)
\label{eq:deltanp1}
\end{equation}
where $\mathbf{J}_F$ is the Jacobian defined by
\begin{equation}
\mathbf{J}_F =\left( \frac{\partial F_i}{\partial \delta_k}  \right )=\delta_{ik}-\mathbf{A}. \varphi'(\delta_k)
\label{eq:jacobien}
\end{equation}
According to Eq.~(\ref{eq:jacobien})
\begin{equation}
\mathbf{J}_F\left( \delta\right) = \mathbf{I} - \mathbf{Q}
\end{equation}
with
\begin{equation}
\mathbf{Q}=\alpha_{ij}.\varphi'(\delta_k)
\end{equation}
At each computation step, $\mathbf{J}^{-1}_F$ is calculated and numerically inverted in order to compute $\delta_{n+1}$ from $\delta_{n}$. The calculation is repeated until $\sum|\delta_{n+1}$-$\delta_{n}|$ is lower than a prescribed convergency criterion.